\documentclass[12pt,preprint]{aastex}

\def\dexkpc{~dex~kpc$^{-1}${}}

\def\kms{~km~s$^{-1}${}}
\def\kmskpc{~km~s$^{-1}$~kpc$^{-1}${}}

\begin{document}

\title{ON THE GALACTIC DISK METALLICITY DISTRIBUTION \\
FROM OPEN CLUSTERS \\
I. NEW CATALOGUES AND ABUNDANCE GRADIENT }

\author{L. Chen, J.L. Hou AND J.J. Wang}
\affil{Shanghai Astronomical Observatory, CAS, Shanghai 200030,
       China}

\email{chenli@center.shao.ac.cn, hjlyx@center.shao.ac.cn}

\begin{abstract}
In this paper we have compiled two new open cluster catalogues. In
the first one, there are 119 objects with ages, distances and
metallicities available, while in the second one, 144 objects have
both absolute proper motion and radial velocity data, of which 45
clusters also with metallicity data available.

Taking advantages of the large number of objects included in our
sample, we present an iron radial gradient of about
$-$0.063$\pm$0.008 \dexkpc \ from the first sample, which is quite
consistent with the most recent determination of oxygen gradient
by nebulae and young stars, which is about $-$0.07 \dexkpc \ . By
dividing clusters into age groups, we show that iron gradient was
steeper in the past, which is consistent with the recent result
from Galactic planetary nebulae data, and also consistent with the
inside-out galactic disk formation scenarios. Based on the cluster
sample, we also discussed the metallicity distribution, cluster
kinematics and space distribution. A disk age-metallicity relation
could be implied from those properties, although we could not give
conclusive result from the age metallicity diagram based on the
current sample. More observations are needed for metal poor
clusters. From the second catalogue, we have calculated the
velocity components in cylindrical coordinates with respect to the
GSR for 144 open clusters. The velocity dispersion of the older
clusters are larger than that of young clusters, but they are all
much smaller than that of the Galactic thick disk stars.

\end{abstract}

\keywords{open clusters: metallicity and kinematics --- Galaxy: formation
--- Galaxy: evolution}

\section{INTRODUCTION}

Since the seminal work of \citet{els62}, great progress has been
made in understanding the formation and evolution of the Milky Way
galaxy. The progress comes, on one side, from observations
concerning chemical abundances in stars (and clusters), gas clouds
and, on the other side, from improved knowledge relevant to galaxy
formation and evolution.

However, some important quantities related to the chemical
evolution of our Galaxy, such as star formation history, initial
mass function, gas flow, etc., are not yet well understood.
Observational data from the Milky Way disk and halo has put strong
constraints on our understanding of those quantities. Among a
variety of observables, radial abundance gradients along the
Galactic disk is one of the most important constraints on the
Galactic chemical evolution model. The existence of such gradients
is now well established, through radio and optical observations of
HII regions, disk stars, planetary nebulae
\citep{hen99,hou00,chi01,ma02b} and open clusters
\citep{fri95,fri99}. An average gradient of $\sim-$0.06 \dexkpc \
is observed in the Milky Way disk for most of the elements, e.g.
O, S, Ne, Ar and Fe. This magnitude of the observed gradients
constrains the various parameters in the chemical evolution model,
such as the timescales of star formation and infall \citep{pra95}
or any variations of the stellar initial mass function properties
with metallicities \citep{chi01}.

In the last decade, a number of successful models have been
developed related to the chemical evolution of the Milky Way
galaxy, but some important differences exist. One of them concerns
the history of abundance gradient along the Galactic disk: were
they steeper or flatter in the past? Different predictions were
made by various models, although most of them claim that they
could reproduce the majority of the observational properties both
in the solar neighborhood and on the whole disk. Time flattening
evolution is suggested by the models of
\citet{pra95,mol97,all98,boi99,cha99,hou00,cha02}, while the
opposite is supported by models of \citet{tos88,sam97} and
\citet{chi97,chi01}.

The situation is neither settled observationally. Estimated ages
of various types (PNI, PNII, PNIII) of planetary nebulae(PN) span
a large fraction of the age of the Galaxy. Observations of the
abundances of those objects across the Milky Way disk could, in
principle, provide some information on the time evolution of the
abundance gradient \citep{mac94,mac99,ma02a,ma02b}. In a recent
work, \citet{hou00} have made a detailed analysis for the O, Ne, S
and Ar gradient based on the PN data of \citet{mac99}. It was
shown that there is fairly good agreement between model
predictions and observations concerning all the properties of the
observed abundance profiles (absolute values, gradient, scatter)
for O, S, Ne and Ar. The model suggests that abundance gradients
are steeper in the earlier epoch. However, the large scatter in
the adopted data does not allow one to conclude on the temporal
variation of the gradients. Nevertheless PNs suffer from large
uncertainties concerning their progenitor's masses and lifetimes
as well as their distances from Galactic center.

On the other hand, open clusters(OCs) have long been used to trace
the structure and evolution of the Galactic disk \citep{fri95}.
Since open clusters could be relatively accurately dated and we
can see them to large distance, their [Fe/H] values serve an
excellent tracer to the abundance gradient along the Galactic disk
as well as many other important disk properties, such as
Age-Metallicity Relation(AMR), abundance gradient evolution, disk
age and so on \citep{car98}.

At this point, one might ask whether the field disk populations
are also able to trace the disk evolution. Indeed, the extensive
studies by \citet{edv93}, and recently by \citet{che00}, who
concentrate on disk F, G stars, show an overall radial gradient
that is nearly independent of age. Those results are based on
stars mainly restricted in the solar neighborhood. A more detailed
analysis for the disk iron gradient was given by \citet{cui00} on
the basis of 1302 field star with high resolution proper motion
and parallax data from Hipparcos satellite. They have derived an
radial iron gradient of $-$0.057 \dexkpc \ within galactocentric
distance from 8.5 kpc to 17 kpc. However, it is still difficult to
reveal any pronounced gradient evolution from those results.
Moreover, results from those studies are strongly affected by
selection effects and rely on the techniques for determining
individual stellar distances (which are heavily dependent on the
adopted Galaxy potential model) that are much less reliable than
those used to obtain cluster distances. In a recent work,
\citet{cor01} have modelled the effects of the orbital diffusion
of stars and clusters on the Galactic abundance gradient. The
general conclusion is that the effect of diffusion makes a
gradient shallower over time, and the cluster population offers a
more viable means for finding detailed structure within the recent
Galactic abundance gradient.

Here, we also pointed out that our recent treatment on deriving
the abundance gradient from open clusters in \citet{hou02} is in
fact not proper. In that paper, we have simply taken four Catalogs
from literatures \citep{car98, twa97, pia95, fri95}, and merge
them just by making cross checking for the common clusters,
without examining individually to see if there are important
difference among clusters in the different catalogs \citep{twa02}.
We refer this paper to act as a substitution to our previous one.

In this paper, we compiled a set of new open cluster catalogues.
The catalogue was divided into two parts: CAT 1 and CAT 2. In CAT
1, we list 119 clusters with iron abundance, age, distance and
reddening data available. This could provide statistically more
significant information to the Galactic disk formation and
evolution, such as Age-Metallicity Relation, abundance gradient
and its time or/and spatial evolution and so on. The second sample
consists of 144 clusters with three dimensional kinematics
information available. From this sample, we are able to explore
some statistical relations among kinematics and other observables.

The paper is organized as the following: firstly in Section 2, we
describe the main characteristic of the two samples. Then, in
Section 3, we give some statistical analysis for the sample,
mainly some metallicity and kinematics distributions. The
abundance gradient is given in Section 4. We show that, based on
our open cluster data, the abundance gradient of the Galactic disk
was steeper in the past. In Section 5, we make some detailed
discussions about the Age-Metallicity Relation (AMR) of the disk.
Finally, a brief summary is given in Section 6.

\section{THE CATALOGUE }

During the past decades, a number of authors have presented their
statistical studies on the Galactic disk based on their own open
cluster catalogue. However, most of the catalogues suffer from
either lacking of homogeneity in the cluster age and metallicity
or insufficient three dimensional kinematic data.

With the full release of Hipparcos Catalogue \citep{esa97} and the
latest Tycho 2 catalogue \citep{hog00}, we have seen a large
growth of proper motion data for open clusters (e.g.
\citet{bau00,dia01}). A most recent compilation was given by
\citet{dia02}. In their catalogue, information of 1537 open
clusters were present, from which $9\%$ have both mean proper
motion and radial velocity data simultaneously; $37\%$ have
distance, $E(B-V)$ and age determinations, including 96 clusters
also have iron abundances data available.

We have compiled two new catalogues of the Galactic open clusters.
The first one (hereafter CAT 1) lists 119 (including Berk29)
clusters parameters for distance, age and metallicity. The age,
distance and reddening information are all (except NGC1348,
NGC2158 and Tombaugh 2) from \citet{dia02}, while  most iron
abundance data (96 clusters) were taken from \citet{dia02}. The
metallicities of another 23 clusters are from other 10 literatures
\citep{cam85, kub92, fri93, fri95, edv95, pia95, bro96, gra00,
ann02, car02}.

Thus far CAT 1 provides a most complete open cluster sample
concerning the iron abundance, distance and age parameters
together. This sample could provide statistically more significant
information concerning the Galactic AMR, radial iron gradient as
well as its evolution, etc.

In the second catalogue (hereafter CAT 2), we have listed observed
kinematical data from literature for 144 clusters, {\it with both
radial velocity and mean proper motion available}. The mean radial
velocity data are mostly (122 of 144 objects) from a compilation
in WEBDA database
(\url{http://obswww.unige.ch/webda/meanvr.html}), primarily based
on the work of \citet{ras99}. The absolute proper motion of 125
clusters, based on the Hipparcos system, are from \citet{bau00}.
Mean proper motions of additional 16 cluster were added from
compilation of \citet{dia01}, with cluster membership probability
derived by Tycho2 proper motions. Data of NGC2355 comes from
\citet{sou00}, and data of Coma Ber and Pleiade clusters are from
\citet{rob99}. In fact, the above observed kinematic information
constitutes a sub-catalogue of that of \citet{dia02}. But, here in
CAT 2 we have further calculated the three dimensional velocity of
open clusters by combing with radial velocity and mean absolute
proper motion data and give, for each cluster, the velocity
components($\Pi,\Theta,W$) in cylindrical coordinates with respect
to the Galactic Standard Reference (GSR). In addition, for each
cluster, age and iron abundance data are also listed whenever
available. (Notice that in the spatial velocity calculation,
following parameters are adopted for the Sun: galactocentric
distance 8.5 kpc, velocity components relative to LSR (10.0, 15.0,
8.0 \kms \ ) and the rotation velocity : 225.0 \kms \ .

Our catalogue files (Table 1, Table 2) are self-explanatory. CAT 1
(Tab.1) consists of data for 119 (118+Berk29) open clusters. For
each cluster, we list its heliocentric galactic coordinates in
B1950.0 and the following parameters, when available:
galactocentric distance; distance from the Sun; color excess
$E(B-V)$, age in Gyr; mean metallicity, and two reference codes
separated by a coma, where the first number is for age, distance
and $E(B-V)$ data while the second number for iron abundance
value. In CAT 2 (tab.2), we provide information of spatial motion
for 144 clusters. For each object, the following data are listed:
galactic coordinates in B1950.0; mean radial velocity ; mean
proper motions; velocity components in cylindrical coordinates
with respect to the GSR; mean spatial velocity and error; age and
iron abundance when available.

In order to check if the data in our CAT 1 has any significant
systematic difference with other published catalogues, we have
made a comparison with Friel's catalogue \citep{fri95}, with 41
clusters in common. We found that the average difference in
metallicity is less than 0.10dex, well within the typical
observational uncertainty. The average difference for $R_{GC}$ is
about 0.5kpc. Note that the age indicator in Friel's work is based
on the MAI (Morphological Age Index), which was only intended to
provide a relative age ranking of clusters, therefore, it is not
fully comparable. But there still has good overall correlation
between Friel's catalogue and ours.

Our following analysis will be mainly based on those two
catalogues, but excluding cluster Berkeley 29. In CAT 1, Berkeley
29 has the galactocentric radius of 23 kpc, the $E(B-V)$ of 0.15
and a metallicity of -0.18 dex, from the compilation of
\citet{dia02}. However, different values for these parameters of
Berkeley 29 were published in the literature. \citet{kal94} gave a
much smaller galactocentric distance of about 19 kpc, the
reddening $E(B-V)$ larger than 0.21 and based on the CMD
morphology and comparison to other old clusters, he also deduced a
[Fe/H] value of lower than -1. In the work of \citet{nor97}, who
applied a new technique for simultaneous determination of [Fe/H]
and $E(B-V)$, a [Fe/H]=-0.30 and a $E(B-V)=0.01$ were given for
Berkeley 29. As the properties of Berkeley 29 are quite uncertain
we do not include this object in the following calculations. In
general, the uncertainty for the metallicity determinations in
open clusters is about 0.1dex.

\section{STATISTICAL PROPERTIES}

\subsection{The Galactic distribution of open clusters}

Using data from \citet{dia02} for 571 open clusters with distance
and age data, we plotted the cluster positions on an (X,Y)
coordinate system, with the zero point in X at the galactic center
(the Sun is assumed to be at 8.5 kpc) as Figure 1 shows. Here the
full line arc represents the solar circle about the galactic
center. One will find from this figure that in the galactic plane,
young clusters (with ages younger than that of Hyades, 0.8 Gyr,
see \citet{phe94}) distributed quite uniformly around the Sun,
while roughly only $20\%$ of the old clusters are inside the solar
circle, most of the old ones are located further away from the
galactic center than the Sun. This result is quite consistent with
the early comprehensive study of \citet{phe94}. The deficiency of
older clusters in the inner part of the disk has been ascribed to
the preferential destruction of the clusters when they encountered
with giant molecular clouds, which were primarily found in the
inner Galaxy.

The distribution of either old or young clusters perpendicular to
the galactic plane could be fitted by a simple exponential law,
which are plotted in Figure 2. The younger clusters are
distributed on the galactic plane almost symmetrically about the
Sun, with a perpendicular scale height of approximately 57 pc. In
contrast, about $80\%$ old clusters are in the outer disk, outside
$R_{gc}=$10 kpc; this population has a scale height of about 354
pc. The derived scale heights are in excellent agreement with the
early results of \citet{jan88}, \citet{jan94}. \citet{jan94}
divided the open clusters into young and old components according
to cluster's MAI, and derived their scale heights as 55pc and
375pc, respectively. The remarkable agreement shows that the MAI
could really be a good age indicator for open clusters.

The main advantage of our CAT 2 is that we have both radial
velocity and mean proper motion available for 144 clusters. This
provides a chance to probe the velocity projection on the
Galactic plane, as shown in Figure 3. Obviously, most of the
clusters are located in the solar neighborhood and have the
velocity vectors well follow the Galactic rotational pattern.

\subsection{The metallicity distribution}

The metallicity distribution of 118 open clusters is plotted in
the upper panel of Figure 4. Here the iron abundance of about 3/4
of the $OC$ sample (with [Fe/H] $>$ -0.2) has roughly a Gaussian
distribution which peaks at the solar value. Meanwhile a metal
poor metallicity tail is also clearly seen. Here we divide our
$OC$ sample by two groups, that is, Metal-Poor (hereafter MP)
component and Metal-Rich (hereafter MR) component, with the
dividing line between them, somewhat arbitrarily, at
[Fe/H]$=-$0.2. In the lower panel of Figure 4, we show two
histograms for the open clusters of the above two groups. By
assuming an exponential law, we can derive their scale-heights to
be 535 pc and 106 pc, for the MP and MR components, respectively.
Taking the scale heights of the Galactic thick and thin disk as
760 pc and 260 pc \citep{ojh96}, we could see that spatially, the
MP group might be either within the tail of the thick disk or in
the outskirts of the thin disk while the MR group is just a thin
disk component. However, our $OC$ metallicity sample is surely not
a complete one, and is subject to a variety of observation
effects. For example, the outer disk clusters are subject to
significant selection effects - they can be seen more readily if
they are at larger distances from the plane, and they are likely
to be older, since younger clusters which live closer to the plane
will not be as visible, and more difficult to observe.

This can also be seen clearly in Fig.5, where we have plotted the
dependence of clusters vertical height on the galactocentric
distance for the whole sample with distances data available
(Fig.5a) and for those also have metallicity data (Fig.5b). In
Fig. 5b, we see that most of the MP $OCs$ are outer disk objects
(with a median $R_{gc}\sim$ 11.3 kpc and the majority(65\%) are
older than 0.8 Gyr, see also Fig. 8,9) and relatively far away
from the Galactic plane, with a median distance about 326 pc. For
MR clusters , most of them are in the inner disk ($R_{gc}\le$10
kpc), distributing in the immediate solar neighborhood (with a
median $R_{gc} \sim$ 8.7 kpc), and about 78\% are young (with ages
$<$ 0.8 Gyr) objects (also see Fig. 8,9). This phenomena is very
likely to imply the possible existence of the age metallicity
relation in open clusters.

Meanwhile, when $R_{gc} < $10 kpc, most $OCs$, no matter MP or MR,
are much closer to the Galactic plane, with a median height z
$\sim$ 84 pc. Especially, there is few clusters observed in the
region of $R_{GC} < $6 kpc. This has been attributed to the
destructive power of the large numbers of giant molecular clouds
in the inner regions of the Galaxy \citep{van80}. There is much
evidence that leads us to believe that open clusters have been
selectively destroyed near the plane of the disk and only those
clusters whose orbits keep them away from the Galactic plane can
survive long enough to appear as outer disk, or metal-poor
clusters. In the meanwhile, this also imply that part of the outer
clusters (they are metal-poor and with high-z) might be formed as
a result of disturbances to the Galactic disk, possibly caused by
tidal interactions with other galaxies or infalling gas, as
suggested by \citet{jan94}. However, it is still hard to
understand why there are almost no high $z$ clusters in the region
of $R_{GC} \sim$ 6.5-8 kpc, compared with the outer disk results.

As a comparison, both globular and open cluster metallicity
distribution are plotted in the upper and lower panels of Figure
6. We can see a clear overlap between metal rich globular cluster
and metal poor open clusters around [Fe/H]$\sim$ $-$0.4 dex. If
the age metallicity relation do exist, then this could be another
evidence which support the idea that a possible connection between
the halo and disk population exist, both in their chemical and
dynamical history.

Based on a survey of proper motion stars, \citet{car90} pointed
out that the Galactic halo population had the chemical and
dynamical history almost independent of the disk. However, from a
study of the oldest open clusters, \citet{phe94} found that the
oldest open clusters (Be17 with an age about 12.5Gyr) have the
ages compatible with that of the youngest globular clusters,
suggesting that there may have been little or no delay in time
between formation of the halo and the onset of the development of
the disk. Phelps's argument rests largely on the age of Be17
clusters, however, recent works have given an age of about 9 Gyr
to this clusters \citep{car99}, and so there does still appear to
be a gap between the formation of the halo and the thick disk. On
the other hand, if the cluster metallicity is related to their
age, the metallicity overlapping of the clusters, as we have
presented, may be another indication of the connection proposed by
\citet{phe94}. In fact, the distinction between "open" and
"globular" (so-called "super") clusters may turn out to be largely
an artificial one \citep{lar02}. They could be both originated
from the Super Star Clusters (SSCs) which are observed in large
numbers in interacting galaxies and merger remnants. Our Milky Way
disk is very likely undergone a process of minor mergers in the
early epoch. The thick disk is plausibly the result of heating of
the thin disk through such events \citep{wys01}.

\subsection{Metallicity vs. kinematic}

In Figure 7, we plot the relation between cluster rotational
velocities around the Galactic center, $\Theta$, and the
galactocentric distance. There exhibits,from least-square fitting,
an insignificant slope of about -2.5 \kmskpc, with quite large
scatter. In the right panel, we present the dependence of velocity
dispersion on the cluster age. The young clusters have a smaller
velocity dispersion, which is expected from their small z scale
height. Although the velocity dispersion for the older clusters is
about 20 \kms \ , it is still much smaller than that of the
Galactic thick disk stars, which is about 50 \kms \ . Anyway, for
most of the CAT.2 clusters (~90$\%$), their heights from the
galactic plane are well within 200pc, they are just thin disk
objects.

In our kinematic sample, there are only 2 objects in the outer
disk (Berkeley 31 and Dolidze 25, both with $R_{gc}>$14 kpc). They
have unreliable proper motion and radial velocity results (with
relative errors up to about 50\%) and thus were not included in
the above radial gradient fitting.

\section{DISK METALLICITY GRADIENT FROM OPEN CLUSTERS}

\subsection{The abundance gradients }

The first radial metallicity gradient using open clusters was
given by \citet{jan79} based on DDO and UBV photometric data of 41
disk objects (part of them are field stars). The derived gradient
is $-$0.05 \dexkpc. \citet{pan81}, by matching theoretical
isochrones to HR diagrams of 20 clusters with age less then 1 Gyr,
derived an iron abundance gradient of $-$0.095 \dexkpc. A similar
result was also obtained by \citet{cam85} based on 37 clusters
with mixed ages. By introducing a weighting system in order to
evaluate and compare the published parameters in \citet{Lyn87}
Catalog of Open Clusters data, \citet{jan88} determined some basic
parameters of 413 open clusters, such as ages, distances, linear
diameters and so on. Among them, 87 open clusters have the
metallicity data. They have derived a gradient about $-$0.133
\dexkpc. By separating the clusters into age groups, they found
that young clusters has much smaller gradient than that of older
clusters. Besides, all those authors found some indications that
the gradient became shallower in the direction of the Galactic
center and steepen in the outer parts of the Galaxy.

\citet{fri93}, (hereafter FJ93) presented their results from a
spectroscopic study of a sample of giant stars in 24 open
clusters.  They derived a galactocentric radial abundance gradient
of [Fe/H] about $-$0.088 \dexkpc. A subsequent revision of the
FJ93 result was presented by \citet{fri95}, using additional
spectroscopic results and a more uniform set of cluster
properties. From a sample of 44 clusters, \citet{fri95} derived an
iron gradient of $-$0.091 \dexkpc. At the same time, \citet{pia95}
derived a much smaller gradients, $-$0.07 \dexkpc, from a sample
of 63 open clusters with a wide range of ages.  These results are
quite consistent with the recent result of \citet{fri99} who
obtained a gradient about $-$0.06 \dexkpc \ . Another gradient
result was presented by \citet{car98} recently. The metallicities
of all selected 37 clusters were obtained spectroscopically. The
final gradient was about $-$0.085 \dexkpc, agreed with earlier
result of FJ93. By dividing the sample into age bins, it was found
that the present-day gradient is a little shallower than the past
one, while the middle epoch seems to display a steepening of the
gradient.

The presence of a linear gradient for open clusters has been
questioned by \citet{twa97} (hereafter TAA97). TAA97 put forth an
alternative description, namely, step function, about the radial
abundances distribution of the open clusters. Within this work, a
set of 76 clusters with abundances based upon DDO and/or moderate
dispersion spectroscopy has been transformed to a common
metallicity scale and used to study the local structure and
evolution of the Galactic disk. They found that the metallicity
distribution of clusters with galactocentric distance is best
described by two distinct zones, with a sharp discontinuity at
R$_{GC}$ = 10 kpc. Between R$_{GC}$ = 6.5 kpc and 10 kpc, the
clusters have a mean metallicity of 0.0 dex with, at best, weak
evidence for a shallow gradient over this range, while those
beyond 10 kpc have a mean value about $-$0.30 dex. This two-step
distribution seems quite similar with the nebula results of
\citet{sim95}. Neglecting this two-step phenomena, a least square
fitting results in a gradient about $-$0.067 \dexkpc between 6 and
15 kpc if cluster BE21 was excluded because both metallicity and
distance of this object are quite uncertain.

The existence of radial iron abundance gradients is also confirmed
by our new up-to-date sample. The result is shown in the upper
panel of Figure 8. By equal-weighted least-square fitting, we
derived a radial abundance gradient of $-$0.063 $\pm$ 0.008
\dexkpc \ , which agrees well with most of the previous open
cluster results. And it is also similar to the gradients obtained
from other tracers, such as disk HII regions and planetary nebulae
(see a summary in \citet{hou00} ). The existence of gradient along
the galactic disk provides good opportunity to test theories of
disk evolution and stellar nucleosynthesis. It suggests that the
role of the Galactic bar in inducing large scale radial mixing and
therefore flattening the gradient has been rather limited;
alternatively, the bar could be too young($<$1Gyr) to have brought
any important modifications to the gaseous and abundance profile.
However, we must notice that our current knowledge on the iron
gradient as derived from open clusters is far from being clear.
Open clusters span a wide range of age, from several millions
years to several Gyr, therefore they do not trace the young
component of the galactic disk. The result we obtained is somewhat
an averaged one (over age). The obtained similarity of gradient
between iron and other elements, such as oxygen, is quite
surprising since the sites of nucleosynthesis for iron and oxygen
are quite different. It is well know that iron is mainly produced
in type SNIa, while oxygen is largely a product of SNII, that is
from massive stars. So the abundance history is very different for
those two types of elements. The gradient similarity might be
simply coincidental, or further investigations should be given in
the production nature of those elements.

We also derived a vertical abundance gradient of $-$0.295 $\pm$
0.050 \dexkpc (lower panel of Figure 8). This is consistent with
the result of \citet{car98}.

\subsection{Gradient evolution in the Galactic disk}

As we have pointed out in Section 1, the behavior of gradient
evolution along the Galactic disk is a major problem for different
chemical evolution models. Open cluster system is an ideal
template for this analysis because $OCs$ have relatively well
determined ages, distances and metallicities.

In the upper panel of Figure 9, we show gradients for two
sub-samples with cluster age $<$ 0.8 Gyr (80 clusters) and $\ge$
0.8 Gyr (38 clusters), respectively. The fitting results are
$-$0.024 $\pm$ 0.012 \dexkpc \ for younger clusters, $-$0.075
$\pm$ 0.013 \dexkpc \ for older ones. If we take the mean age for
the youngest and oldest clusters as 0.00 Gyr and 6.00 Gyr (this is
somewhat arbitrary, just for illustration purpose) in our sample,
we can estimate an average flattening rate of 0.008 \dexkpc
Gyr$^{-1}$ during the past 6 Gyr. Similar value is obtained by
\citet{ma02b} from PN data for [O/H].

As we have indicated in the Introduction, that the time evolution
of the abundance gradient along the Galactic disk is crucial in
discriminating different theoretical models that adopt various
prescriptions used for the time dependence of the SFR and the
infall. Our current open clusters sample could surely provide some
insights on this subject. The time flattening tendency we obtained
supports the 'inside-out' disk formation scenarios with infall
time scale dependent on radius from the disk center
\citep{boi99,hou00,cha02}.

In the lower panel of Figure 9, we divided clusters into inner (
$<$ 10 kpc) and outer groups. The corresponding gradients are
$-$0.040 $\pm$ 0.022 \dexkpc \ and $-$0.047 $\pm$ 0.023 \dexkpc,
respectively. We can see that the inner disk exhibits roughly the
same (or a bit smaller) gradient as the outer part. This result is
also consistent with the abundance gradient determined by using
Cepheid in the solar neighborhood \citep{and02}. However, in our
CAT 1, the inner most cluster is located at a galactocentric
distance about 6.8 kpc, it is necessary to have more inner
clusters data (between 3 kpc and 7 kpc ) in order to further check
the gradient behavior for the inner disk. If the Galactic bar does
play the role, then the inner gradient could be more flat compared
with outer part.

Our cluster sample is nearly 50 $\%$ more than that of TAA97, and
we did not find evidence of any abrupt discontinuity. A similar
conclusion was reached by \citet{fri99}, using high resolution
abundance determinations for metallicity calibration.

\section{DISK AGE-METALLICITY RELATION}

The age-metallicity relation(AMR) for the Galactic disk provides
useful clues about the chemical evolution history of the Milky
Way, and also put an important constraint on the theoretical
models of the disk. The observed abundance data generally show a
decrease of the stellar metallicity with increasing stellar age,
indicating a continuous growth of the metals in the ISM during the
life of the Galaxy. The early study on AMR for nearby stars by
\citet{twa80} found that the mean metallicity of the disk
increased by a factor of five between 12 and 5 billion year ago
and has increased only slightly since then. This was also
confirmed by latter photometric survey of \citet{meu91}. With the
high resolution spectroscopic data, \citet{edv93} showed a plot of
iron abundance versus relative ages for the 189 stars in the solar
neighborhood. The overall trend of a slowly increasing abundance
with decreasing age was consistent with the previous photometric
results. However, the most striking feature of their result is the
large scatter around the average trend, which marks a weak
correlation between age and metallicity. This spread was, as they
pointed out, in part due to selection bias for the programme
stars, and at least partly intrinsic, since the mean errors in
[Fe/H] measurement and logarithmic age derivation are much less
than the scatter.

In a recent paper, \citet{fel01} have re-examined the Galactic AMR
in the solar neighborhood based on a sample of 5828 dwarfs and
sub-dwarfs from Hipparcos Catalogue. They found that the solar
neighborhood age-metallicity diagram is well populated at all ages
and especially that old, metal-rich stars do exist, which have
been omitted in previous samples. This indicates a complete lack
of enrichment over the age of Galactic disk among the fields stars
in the solar neighborhood.

Using open clusters to explore the AMR has the main advantage both
in abundance and age determinations since one is dealing with a
group of stars and the result is less susceptible to individual
errors \citep{car98}. \citet{cam85} was the first to probe the AMR
from open cluster data, and found no age-metallicity relation
based on his cluster sample.  This is not surprising since the
metallicity of the Galactic disk increased only slightly during
the past 5 Gyr, while his sample of 38 clusters contained no
objects older than 5.1 Gyr. More recently, \citet{car98} compiled
a relatively homogenous sample of 37 open clusters. The data have
more expanded cluster ages up to 9 Gyr. After correcting for the
radial abundance gradient, the derived AMR showed similar trend to
that of nearby stars.

In this paper, we have present a new open cluster catalogue with
much more objects. The results, based on this larger sample, would
be statistically more reliable. As we have shown in the Sect 3.2,
statistically, the space distributions (scale heights for metal
poor and metal rich groups, for young and old clusters) of open
clusters are very likely imply the existence of age-metallicity
relation in the Galactic disk. In Fig. 10, we plot the dependence
of metallicity on the cluster age, after correcting the radial
metallicity gradient. Unfortunately, it is difficult to draw any
conclusive indication for AMR based on this plot due to the
deficiency of very older clusters. More observational efforts
should be added in finding more older clusters.

The significant spread of the AMR seems real, but its origin is
not yet clear. For the scatter in the AMR of nearby stars, many
possible causes have been suggested, such as orbital diffusion of
stars, inhomogeneous chemical enrichment in the Galaxy evolution,
overlapping of different galactic substructures and so on. All the
above mentioned effects may contribute the observed scatter, while
for open clusters, the result should not be very sensitive to
orbital diffusion effects \citep{cor01}. Therefore, the scatter of
AMR along the Galactic disk from both clusters and field disk
stars is an essential feature in the formation and evolution of
the Milky Way.

\section{Summary}

The main work of this paper is to compile a most complete open
clusters sample with metallicity, age, distance data as well as
kinematic information available. And upon this sample, some
statistical analysis on spatial and metallicity distributions have
been made.

We derived an iron radial gradient about $-$0.063$\pm$0.008
\dexkpc \ from the CAT 1, which is quite consistent with the most
recent determination of oxygen gradient in nebulae and young
stars. By dividing clusters into age groups, we show that iron
gradient was steeper in the past, which is consistent with the
recent result from Galactic planetary nebulae data. Our result
supports the inside-out Galactic disk formation mechanism that
invoking SFR and infall time scale various with radius.

The spatial variation of gradient was also explored. When the
clusters were divided into inner and outer groups, we found that
the abundance gradient was a bit shallower inside 10 kpc. But the
inner most cluster in our sample is in $R_{GC} = 6.8$kpc, we need
more data of clusters in the inner region. This could be helpful
to judge the radial flow effect on the current galactic chemical
evolution model.

From scale heights of metal poor and metal rich clusters, we
noticed that the metallicity could be related to the age. However,
by plotting directly the dependence of cluster abundancess on
their ages, no strike slope in AMR was found. However, the paucity
of metallicity of very old open clusters made it impossible to
give a definite conclusion based on the current sample.

\acknowledgments

The Authors thank Drs. Chenggang Shu, Ruixiang Chang and Prof.
Chenqi Fu for their helpful discussion. The authors would also
thanks to an anonymous referee for invaluable comments and
stimulating criticism of the manuscript, which greatly improved
this article. This research was supported partly by the National
Natural Science Foundation of China (No.19873014, No.10173017,
No.10133020, No.19833010) and NKBRSFG 1999075404 \& 406, and
partly by SRF for ROCS, SEM.

\def\cjaa{CJAA}
\def\nature{Nature}
\def\apss{Ap\&SS}
\def\aapss{AApSS}
\def\pasp{PASP}
\def\sinc{Science in China}
\def\japa{JAp\&A}

\clearpage

\begin{figure}
\figurenum{1}\plotone{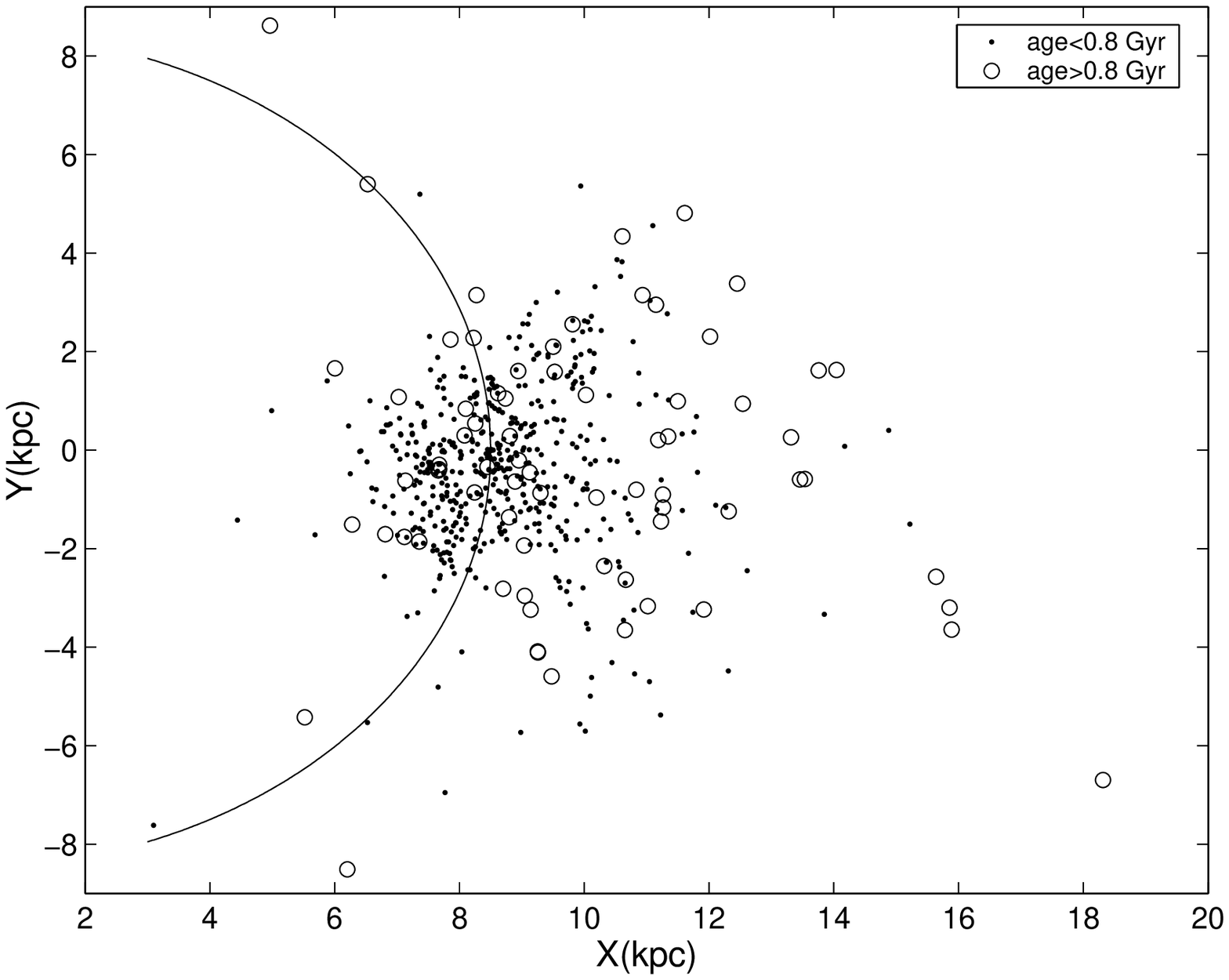} \caption{Spacial distribution
of the open clusters on the galactic plane. The open circles are
old clusters with age great than that of Hyades (0.8 Gyr), and the
dots are for younger ones. The Sun is at X = 8.5,Y = 0 kpc. The
galactic center is at (0,0). The circle has a radius of 8.5 kpc,
centered on the galactic center. \label{fig1}}

\end{figure}

\clearpage

\begin{figure}
\figurenum{2} \plotone{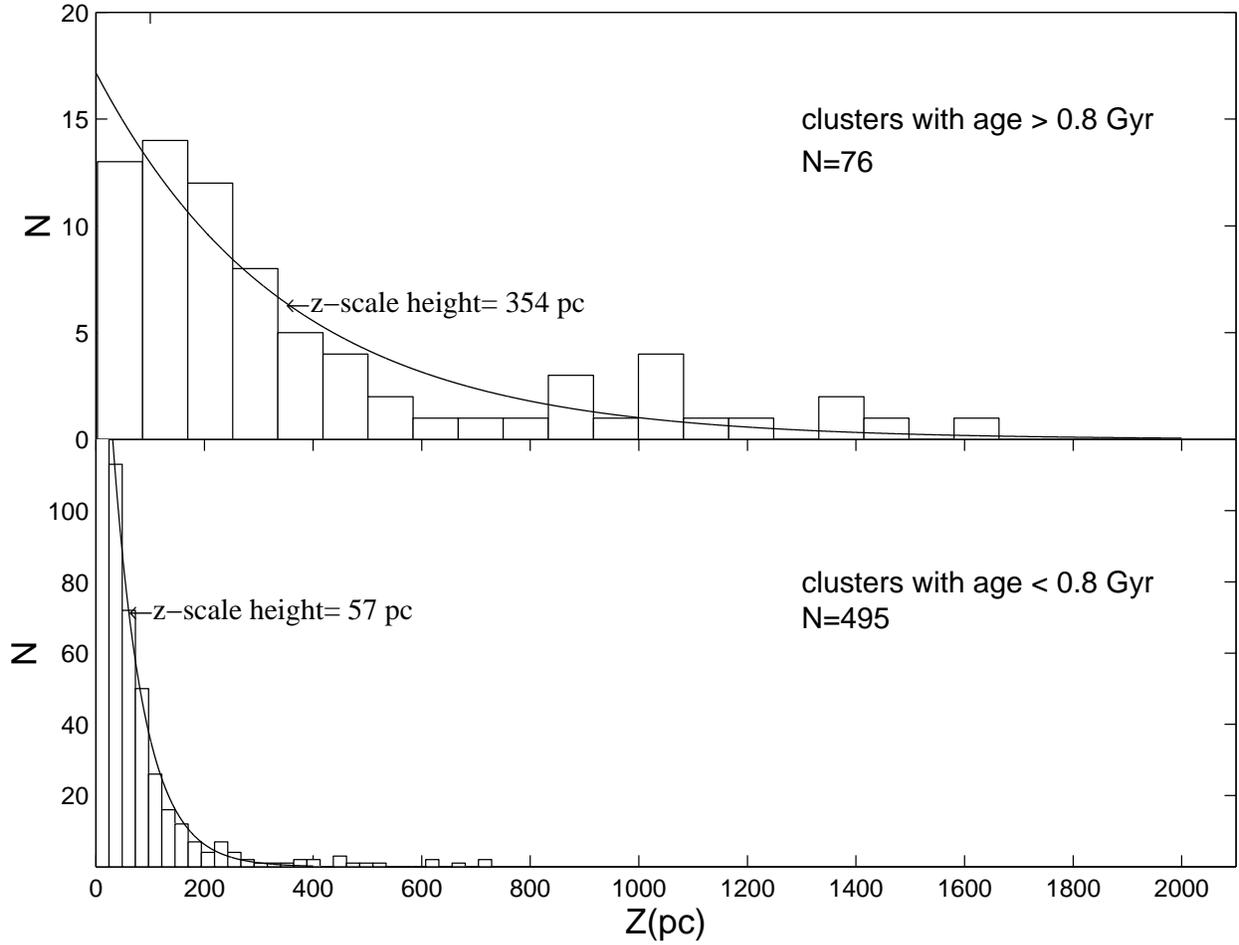} \caption{Number distribution
from the galactic plane, z, for old and young sub-groups of open
clusters. The fitted scale heights for the two groups are 354pc
and 57pc, respectively. \label{fig2}}

\end{figure}

\clearpage

\begin{figure}
\figurenum{3} \plotone{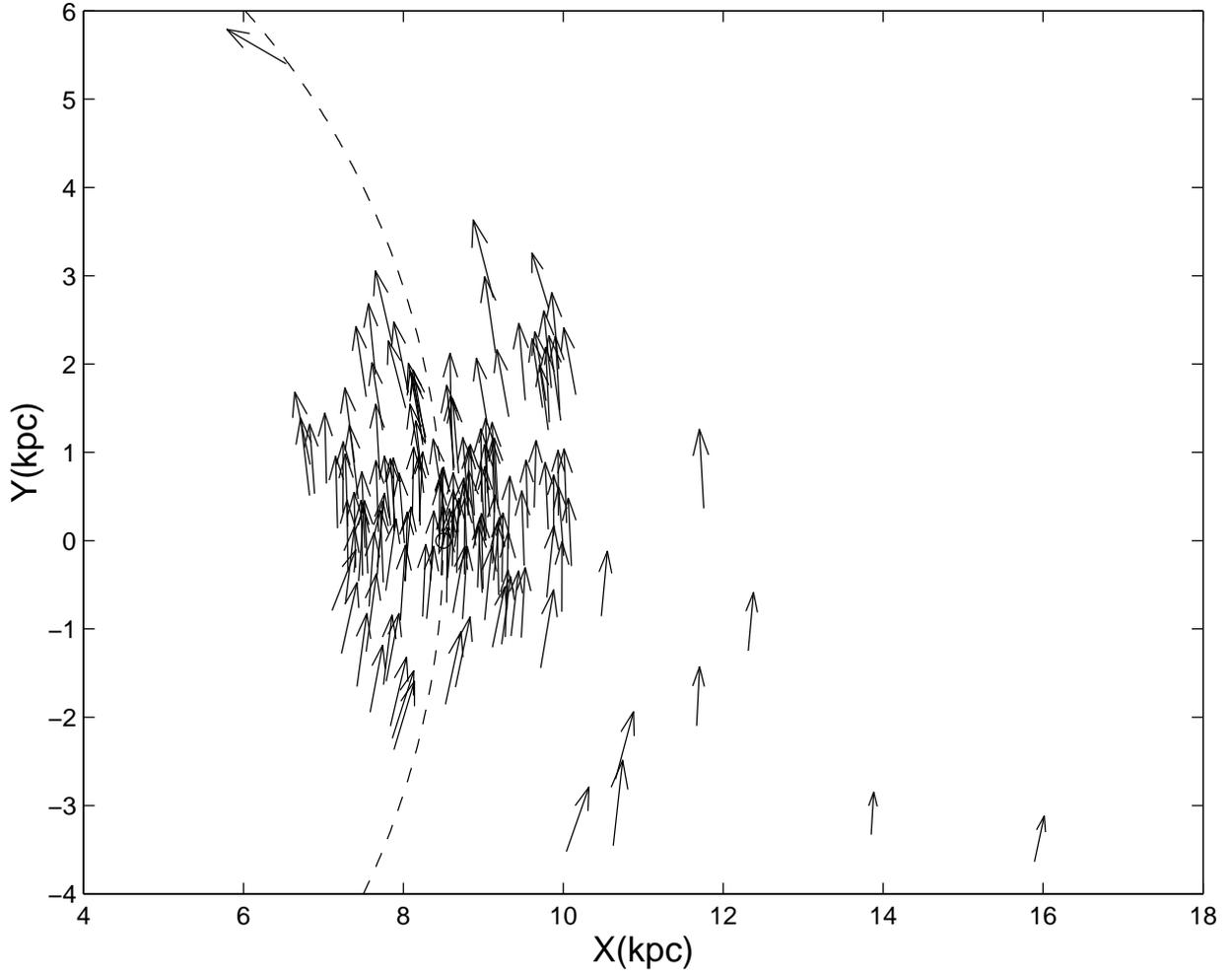} \caption{Velocity projection
on the Galactic plane for 144 clusters, which have both radial
velocity and mean proper motion available. The Sun is located at X
= 8.5 kpc, Y = 0 pc. \label{fig3}}

\end{figure}

\clearpage

\begin{figure}
\figurenum{4} \plotone{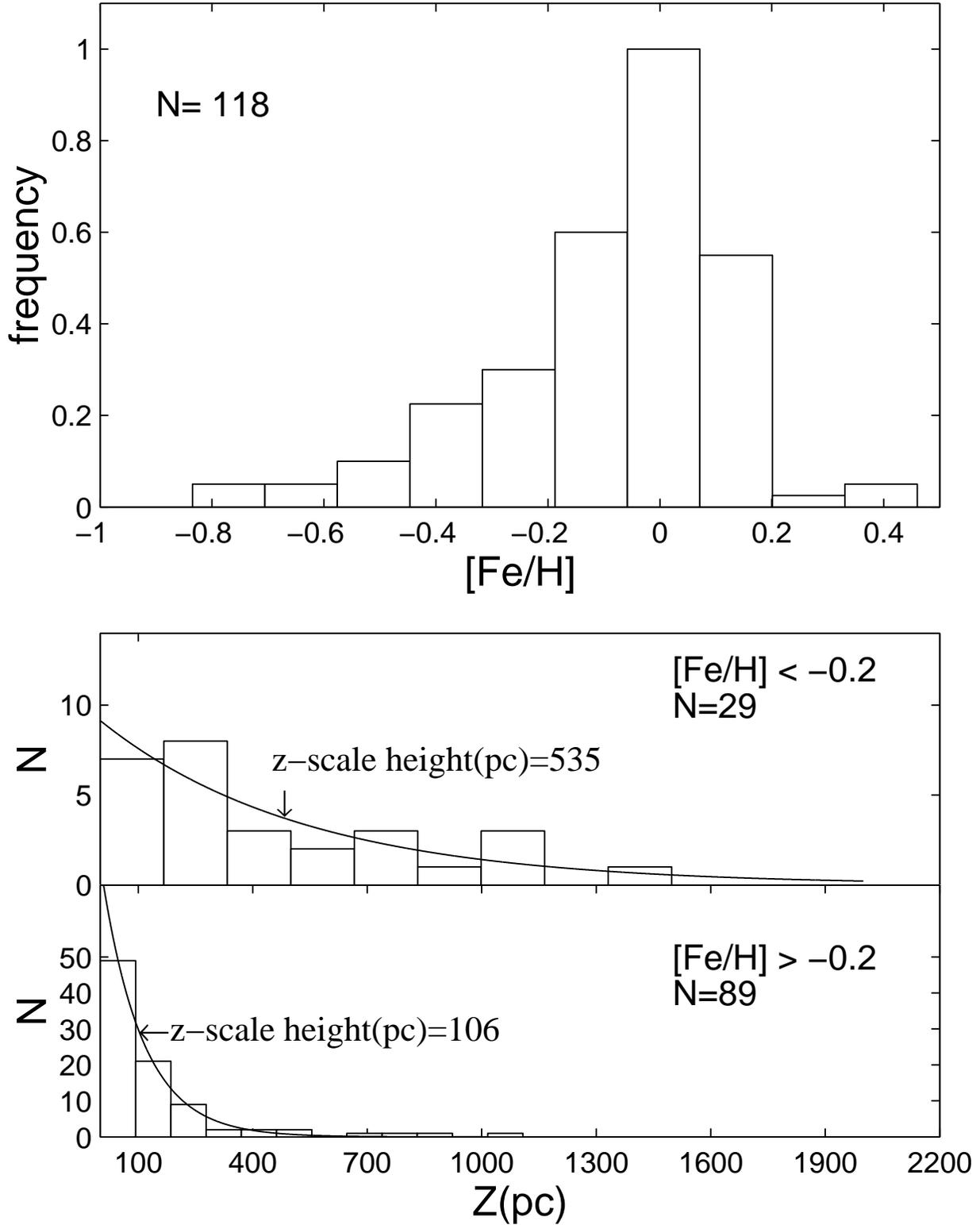} \caption{Upper panel:
metallicity distribution of the Galactic open clusters.
Lower panel: number distribution from the galactic plane, Z, for
metal poor and metal rich sub-groups of open clusters. We can see
that the two groups are different in their scale properties,
with scale height about 535 pc and 106 pc,respectively. \label{fig4}}

\end{figure}

\clearpage

\begin{figure}
\figurenum{5} \plotone{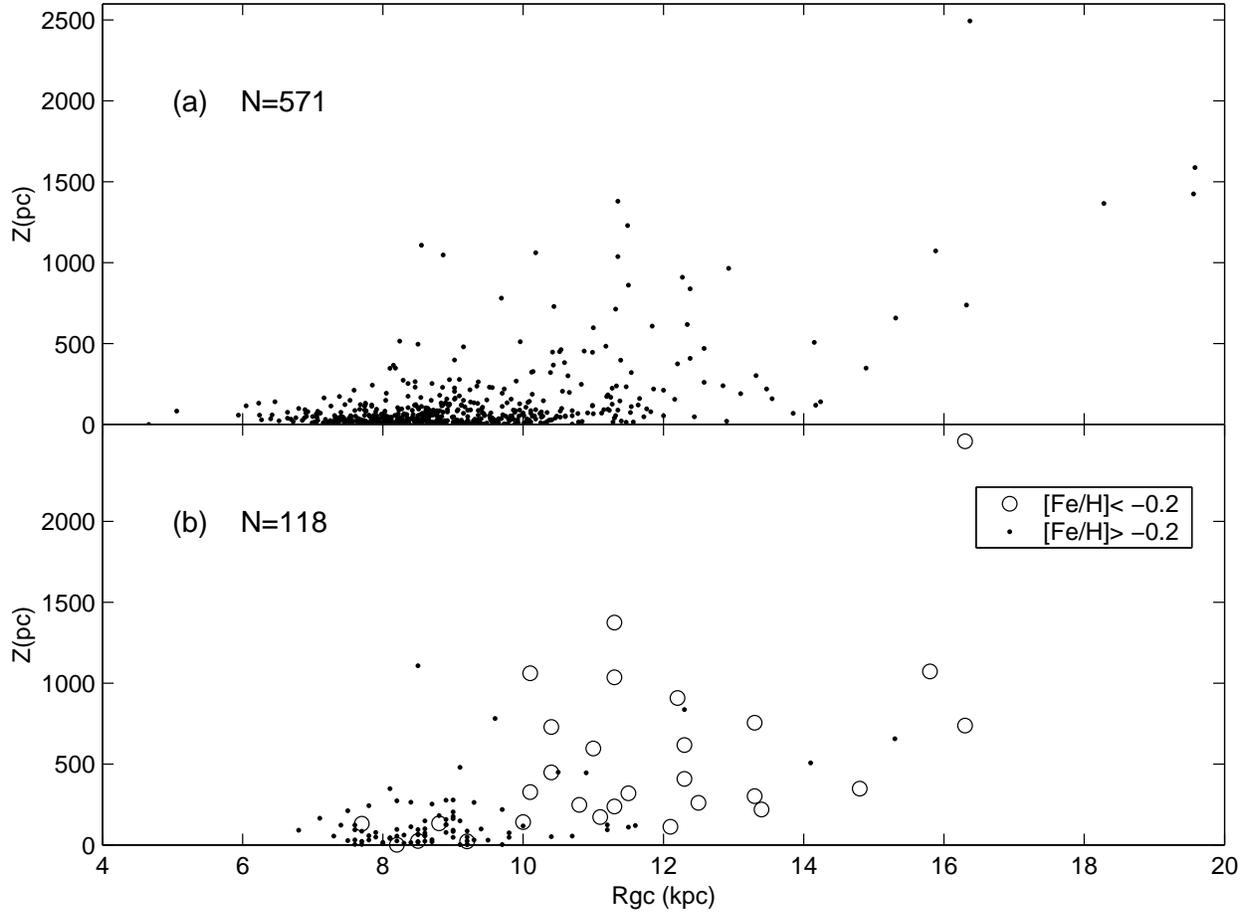} \caption{Upper panel: height
from the galactic plane vs. galactocentric distance, for all 571
clusters with data available. Lower panel: same as the upper
panel, but only for 118 $OCs$ with also iron abundance data. It
can be seen that the most metal poor clusters are located in the
outer part of the Galactic disk.  \label{fig5}}

\end{figure}

\clearpage
\begin{figure}
\figurenum{6} \plotone{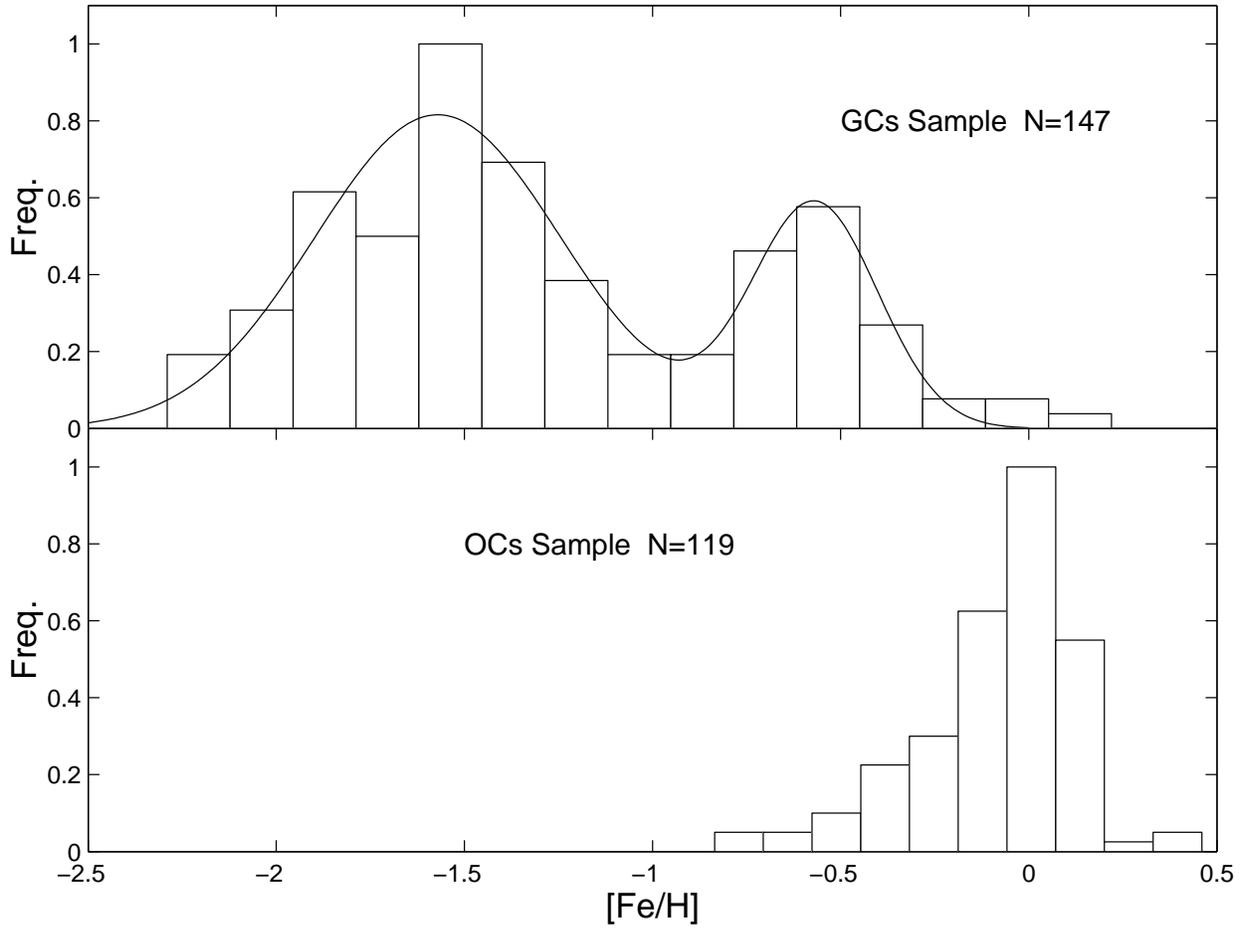} \caption{Comparison of the
metallicity distributions for Galactic globular and open clusters
\label{fig6}}

\end{figure}

\clearpage

\begin{figure}
\figurenum{7} \plotone{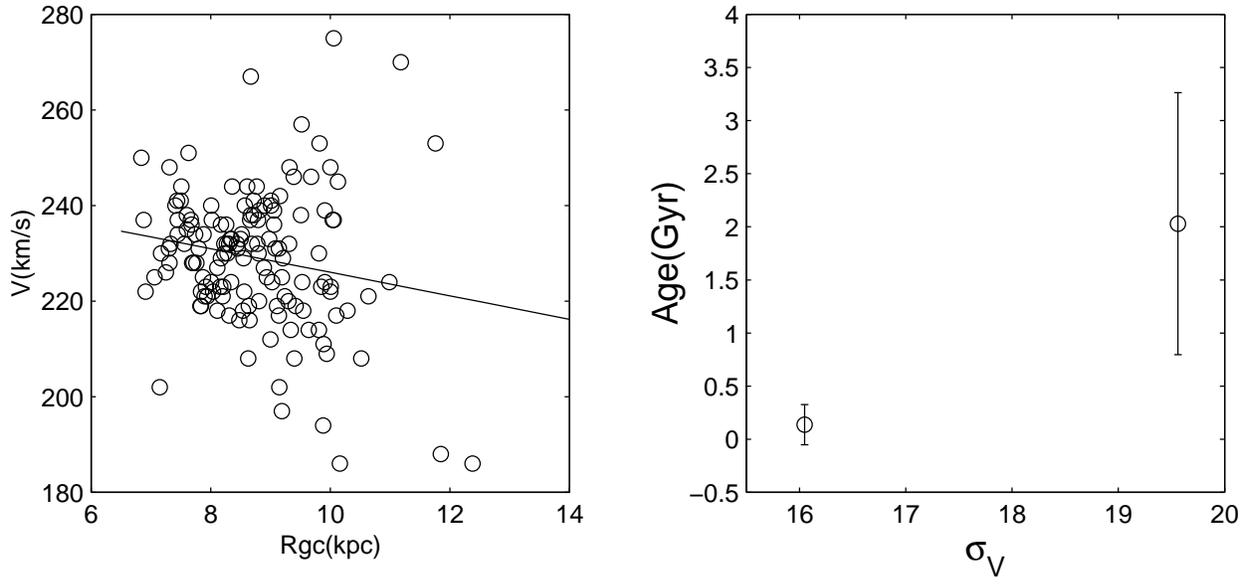} \caption{Left:
correlation between clusters rotational velocity $\Theta$ and
galactocentric distance. Right: velocity dispersion vs. cluster age.
\label{fig7}}

\end{figure}

\clearpage

\begin{figure}
\figurenum{8} \plotone{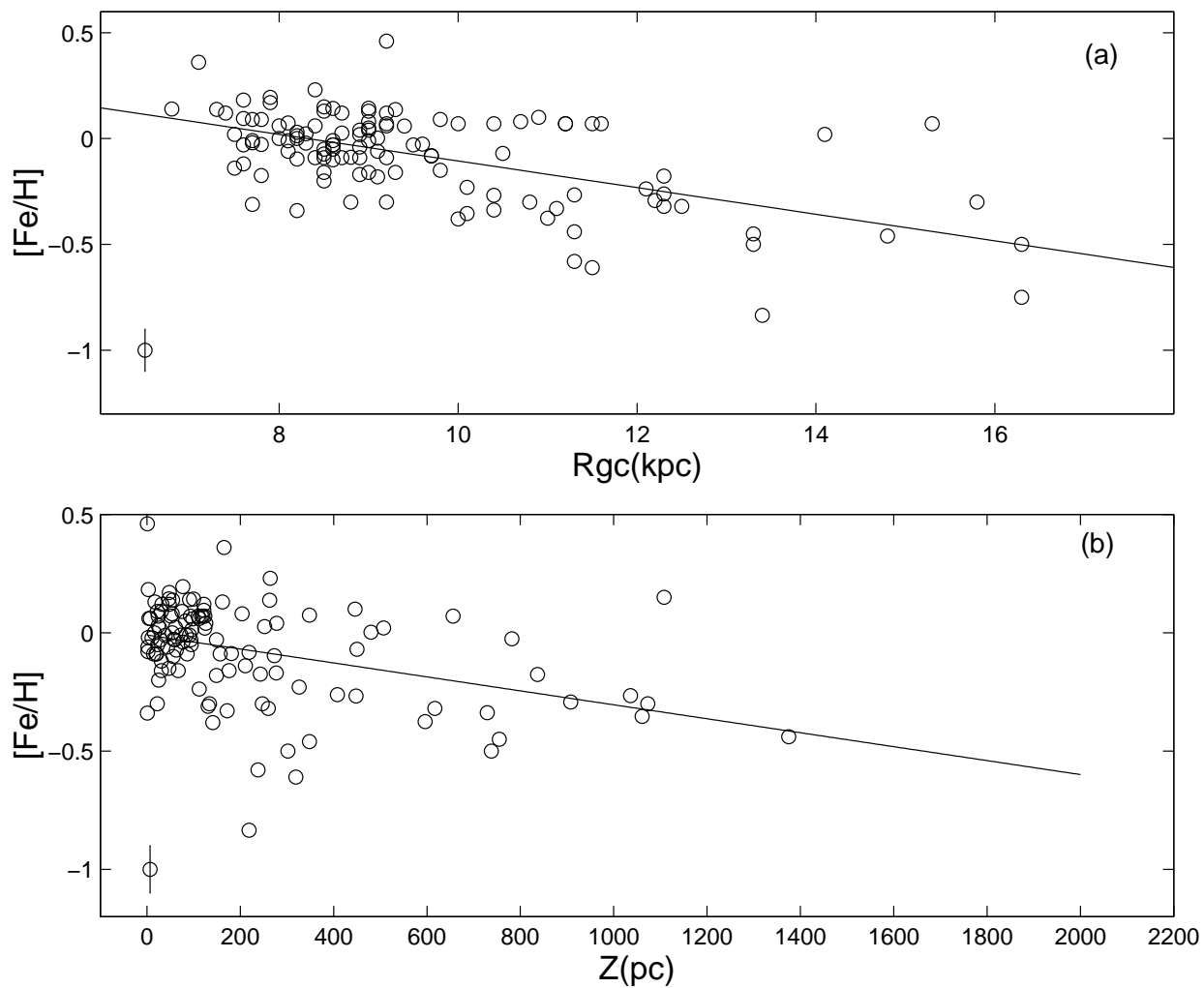} \caption{Radial (upper
panel) and vertical (lower panel) abundance gradient for 118 open
clusters. The least-square fitting results in a gradient of
$-$0.063 $\pm$ 0.008 \dexkpc \ and $-$0.295 $\pm$ 0.050\dexkpc,
respectively. The typical error bar for [Fe/H] is about 0.1dex, as
showed in lower left corner of the figures. When deriving the
vertical gradient, the radial gradient has been corrected.
\label{fig8}}

\end{figure}

\clearpage

\begin{figure}
\figurenum{9} \plotone{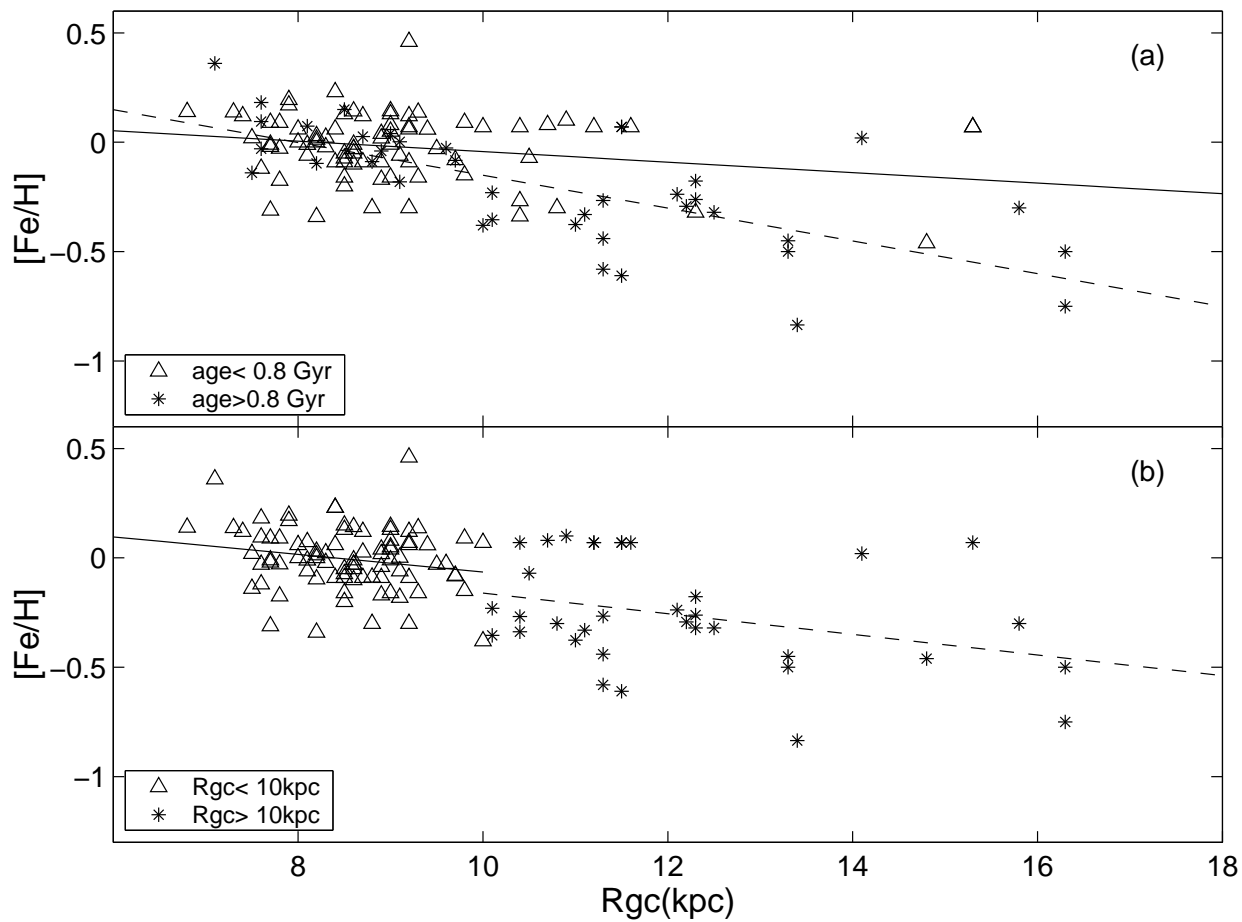} \caption{Upper panel: Time
evolution of the iron gradient. Triangles show clusters with age
less than 0.8 Gyr, stars show clusters with age greater than 0.8
Gyr. The gradients are $-$0.024 \dexkpc \ and $-$0.075 \dexkpc,
respectively. Lower pane: gradients for inner disk (within 10 kpc)
and out disk clusters. The corresponding gradients are $-$0.040
\dexkpc \ and $-$0.047 \dexkpc, respectively. \label{fig9}}

\end{figure}

\clearpage

\begin{figure}
\figurenum{10} \plotone{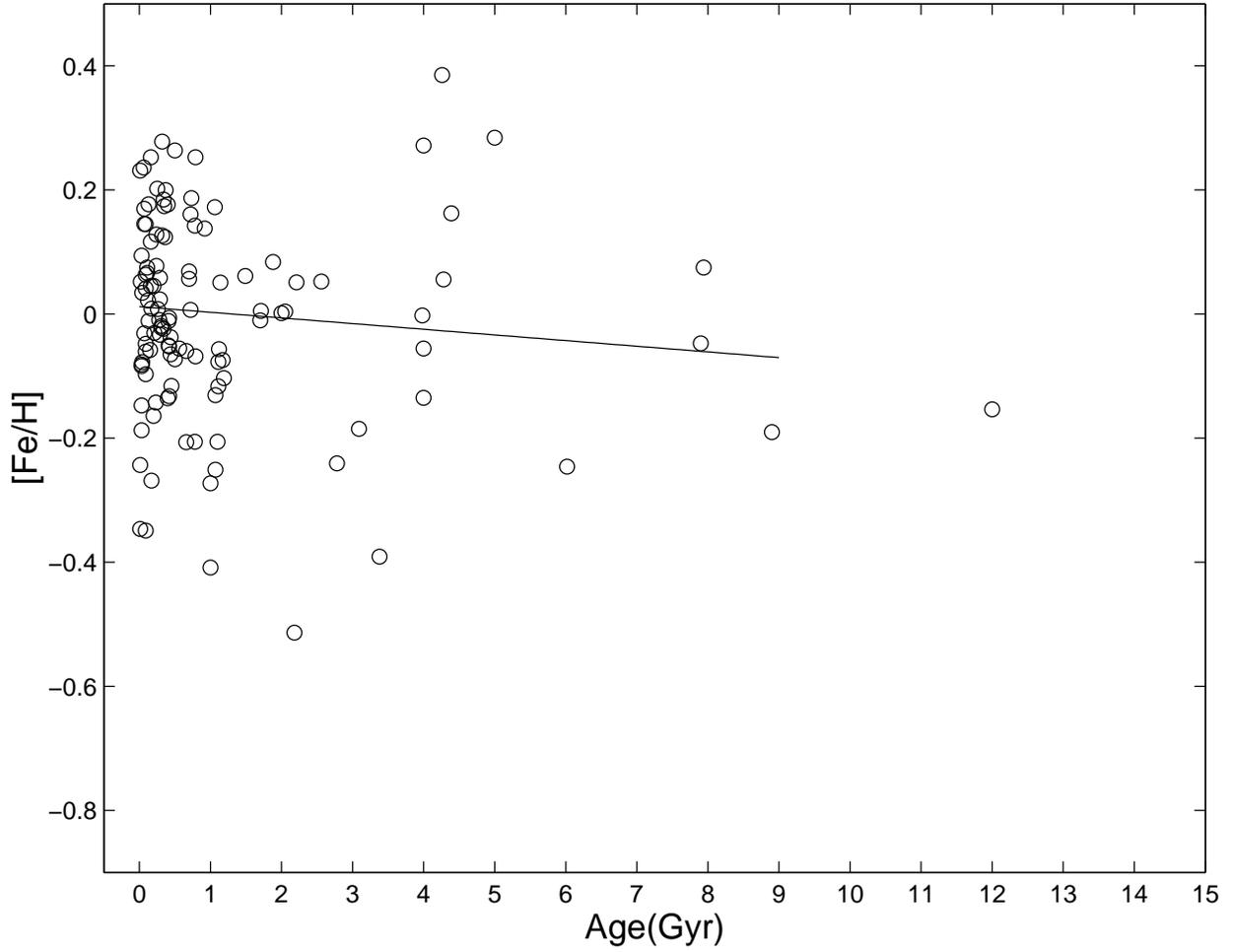} \caption{\label{}
Age-Metallicity Relation (AMR) for the 118 open clusters after
correcting for the radial gradient. The solid line is a
least-square fitting for the open cluster data.
\label{fig10}}
\end{figure}

\clearpage

\begin{deluxetable}{rlrrrrrrrr}
\tablewidth{0pt} \tablecaption{Age,Distance and Metallicity of
Open Clusters} \tablehead{ \colhead{No.} & \colhead{Name}      &
\colhead{l} & \colhead{b}  & \colhead{R$_{GC}$} &
\colhead{R$_{SUN}$} & \colhead{E(B-V)} & \colhead{Age}  &
\colhead{[Fe/H]} &
\colhead{refs}\\
\colhead{ } & \colhead{ } & \colhead{(deg)} & \colhead{(deg)} &
\colhead{(kpc)} & \colhead{(pc)} & \colhead{mag} & \colhead{(Gyr)
} & \colhead{dex} & \colhead{ }
 }
 \startdata
  1 & Berkeley 12   & 161.66 &  -1.99 & 11.5 &  3162 &   0.7 &     4 &  0.07 & 1,1 \\
  2 & Berkeley 17   & 175.65 &  -3.65 & 11.1 &  2700 &   0.7 & 12.00 & -0.33 & 1,1 \\
  3 & Berkeley 18   & 163.63 &   5.01 & 14.1 &  5800 &  0.46 &  4.26 &  0.02 & 1,1 \\
  4 & Berkeley 19   &  176.9 &  -3.59 & 13.3 &  4831 &   0.4 &  3.09 & -0.50 & 1,2 \\
  5 & Berkeley 20   & 203.51 & -17.28 & 16.3 &  8400 &  0.12 &  6.02 & -0.75 & 1,3 \\
  6 & Berkeley 21   & 186.84 &  -2.51 & 13.4 &  5000 &  0.76 &  2.18 & -0.83 & 1,1 \\
  7 & Berkeley 22   &  199.8 &  -8.05 & 15.8 &  7663 &   0.7 &  1.06 & -0.30 & 1,1 \\
  8 & Berkeley 23   &  192.6 &   5.44 & 15.3 &  6918 &   0.4 &  0.79 &  0.07 & 1,1 \\
  9 & Berkeley 29   & 197.98 &   8.02 & 23.0 & 14871 & 0.157 &  1.05 & -0.18 & 1,1 \\
 10 & Berkeley 31   & 206.25 &   5.12 & 16.3 &  8272 &  0.08 &  2.05 & -0.50 & 1,1 \\
 11 & Berkeley 32   & 207.95 &    4.4 & 11.3 &  3100 &  0.16 &  3.38 & -0.58 & 1,1 \\
 12 & Berkeley 39   & 223.46 &  10.09 & 12.3 &  4780 &  0.12 &  7.94 & -0.17 & 1,1 \\
 13 & Berkeley 60   & 118.84 &  -1.63 & 11.2 &  4365 &  0.86 &  0.16 &  0.07 & 1,1 \\
 14 & Berkeley 64   & 131.91 &    4.6 & 11.5 &  3981 &  1.05 &     1 & -0.61 & 1,1 \\
 15 & Berkeley 70   & 166.89 &   3.58 & 12.5 &  4158 &  0.48 &     4 & -0.32 & 1,1 \\
 16 & Berkeley 104  & 117.63 &   1.22 & 11.2 &  4365 &  0.45 &  0.79 &  0.07 & 1,1 \\
 17 & Blanco 1      &  15.57 & -79.26 &  8.4 &   269 &  0.01 &  0.06 &  0.23 & 1,4 \\
 18 & Collinder 140 & 244.97 &  -7.91 &  8.6 &   405 &  0.03 &  0.03 & -0.10 & 1,1 \\
 19 & Collinder 261 & 301.68 &  -5.52 &  7.5 &  2190 &  0.27 &   8.9 & -0.14 & 1,1 \\
 20 & IC 2391       & 270.36 &   -6.9 &  8.5 &   175 & 0.008 &  0.04 & -0.09 & 1,1 \\
 21 & IC 2581       & 284.59 &   0.03 &  8.2 &  2446 & 0.415 &  0.01 & -0.34 & 1,1 \\
 22 & IC 2602       &  289.6 &   -4.9 &  8.4 &   161 & 0.024 &  0.03 & -0.09 & 1,5 \\
 23 & IC 2714       &  292.4 &  -1.79 &  8.1 &  1238 & 0.341 &  0.34 & -0.01 & 1,1 \\
 24 & IC 4651       & 340.08 &   -7.9 &  7.6 &   888 & 0.116 &  1.14 &  0.09 & 1,1 \\
 25 & IC 4725       &   13.7 &  -4.43 &  7.9 &   620 & 0.476 &  0.09 &  0.17 & 1,1 \\
 26 & IC 4756       &  36.38 &   5.24 &  8.1 &   484 & 0.192 &   0.5 & -0.06 & 1,1 \\
 27 & King 5        & 143.74 &  -4.27 & 10.0 &  1900 & \nodata & 1   & -0.38 & 1,1 \\
 28 & King 6        & 143.36 &  -0.07 &  9.2 &   871 &   0.5 &  0.25 &  0.46 & 1,1 \\
 29 & King 8        & 176.39 &   3.12 & 14.8 &  6403 &  0.58 &  0.41 & -0.46 & 1,1 \\
 30 & King 11       & 117.16 &   6.47 & 10.1 &  2892 &  1.27 &  1.11 & -0.23 & 1,1 \\
 31 & King 15       & 120.74 &  -0.92 & 10.4 &  3162 &   0.7 &  0.25 &  0.07 & 1,1 \\
 32 & Melotte 20    & 146.93 &  -7.11 &  8.6 &   185 &  0.09 &  0.07 & -0.05 & 1,5 \\
 33 & Melotte 22    & 166.57 & -23.52 &  8.6 &   150 &  0.03 &  0.13 & -0.03 & 1,5 \\
 34 & Melotte 25    & 180.06 & -22.34 &  8.5 &    45 &  0.01 &  0.78 &  0.13 & 1,5 \\
 35 & Melotte 66    & 259.55 & -14.24 & 10.1 &  4313 & 0.143 &  2.78 & -0.35 & 1,1 \\
 36 & Melotte 71    & 228.96 &   4.49 & 10.8 &  3154 & 0.113 &  0.23 & -0.30 & 1,1 \\
 37 & Melotte 111   & 221.35 &  84.02 &  8.5 &    96 & 0.013 &  0.44 & -0.05 & 1,5 \\
 38 & NGC 1039      & 143.63 &  -15.6 &  8.8 &   499 &  0.07 &  0.17 & -0.30 & 1,1 \\
 39 & NGC 1193      & 146.74 & -12.19 & 12.2 &  4300 &  0.12 &   7.9 & -0.29 & 1,1 \\
 40 & NGC 1245      & 146.64 &  -8.92 & 10.9 &  2876 &   0.3 &   0.5 &  0.10 & 1,1 \\
 41 & NGC 1342      & 154.95 & -15.34 &  9.0 &   665 & 0.319 &  0.45 & -0.16 & 1,1 \\
 42 & NGC 1348      & 146.96 &   -3.7 & 10.0 &  1820 &  1.02 &  0.13 &  0.07 & 6,6 \\
 43 & NGC 1545      & 153.36 &   0.18 &  9.1 &   711 & 0.303 &  0.28 & -0.06 & 1,1 \\
 44 & NGC 1662      & 187.71 & -21.11 &  8.9 &   437 & 0.304 &  0.42 & -0.09 & 1,1 \\
 45 & NGC 1817      & 186.13 & -13.12 & 10.4 &  1972 & 0.334 &   0.4 & -0.26 & 1,1 \\
 46 & NGC 188       & 122.78 &  22.46 &  9.6 &  2047 & 0.082 &  4.28 & -0.02 & 1,1 \\
 47 & NGC 2099      & 177.64 &   3.09 &  9.8 &  1383 & 0.302 &  0.34 &  0.09 & 1,1 \\
 48 & NGC 2141      & 198.08 &  -5.81 & 12.3 &  4033 &  0.25 &  1.7  & -0.26 & 1,1 \\
 49 & NGC 2158      & 186.64 &   1.78 & 12.1 &  3600 &  0.55 &  2.0  & -0.23 & 7,1 \\
 50 & NGC 2168      & 186.59 &   2.19 &  9.3 &   816 & 0.262 &  0.09 & -0.16 & 1,1 \\
 51 & NGC 2204      & 226.01 &  -16.1 & 10.4 &  2629 & 0.085 &  0.78 & -0.33 & 1,1 \\
 52 & NGC 2243      &  239.5 & -17.97 & 11.3 &  4458 & 0.051 &  1.07 & -0.44 & 1,1 \\
 53 & NGC 2251      & 203.58 &    0.1 &  9.7 &  1329 & 0.186 &  0.26 & -0.08 & 1,1 \\
 54 & NGC 2259      & 201.76 &   2.07 & 11.6 &  3311 &  0.59 &  0.32 &  0.07 & 1,1 \\
 55 & NGC 2281      &  174.9 &  16.88 &  9.0 &   558 & 0.063 &  0.35 &  0.13 & 1,1 \\
 56 & NGC 2287      & 231.01 & -10.44 &  8.9 &   693 & 0.027 &  0.24 &  0.04 & 1,1 \\
 57 & NGC 2301      & 212.56 &   0.29 &  9.2 &   872 & 0.028 &  0.16 &  0.06 & 1,1 \\
 58 & NGC 2304      &  197.2 &   8.89 & 12.3 &  3991 &   0.1 &  0.79 & -0.32 & 1,1 \\
 59 & NGC 2335      & 223.62 &  -1.18 &  9.5 &  1417 & 0.393 &  0.16 & -0.03 & 1,1 \\
 60 & NGC 2343      & 224.32 &  -1.17 &  9.2 &  1056 & 0.118 &  0.01 &  -0.3 & 1,1 \\
 61 & NGC 2355      &  203.3 &   11.8 & 10.5 &  2200 &  0.12 &   0.7 & -0.07 & 1,1 \\
 62 & NGC 2360      &  229.8 &  -1.43 &  9.8 &  1887 & 0.111 &  0.56 & -0.15 & 1,1 \\
 63 & NGC 2420      & 198.11 &  19.63 & 11.3 &  3085 & 0.029 &  1.11 & -0.26 & 1,1 \\
 64 & NGC 2423      & 230.48 &   3.54 &  9.0 &   766 & 0.097 &  0.73 &  0.14 & 1,1 \\
 65 & NGC 2437      & 231.87 &   4.07 &  9.4 &  1375 & 0.154 &  0.24 &  0.05 & 1,1 \\
 66 & NGC 2477      & 253.59 &  -5.83 &  8.9 &  1222 & 0.279 &   0.7 &  0.01 & 1,1 \\
 67 & NGC 2482      & 241.63 &   2.03 &  9.2 &  1343 & 0.093 &   0.4 &  0.12 & 1,1 \\
 68 & NGC 2489      & 246.71 &  -0.78 & 10.7 &  3957 & 0.374 &  0.01 &  0.08 & 1,1 \\
 69 & NGC 2506      & 230.57 &   9.92 & 11.0 &  3460 & 0.081 &   1.1 & -0.37 & 1,1 \\
 70 & NGC 2516      & 273.81 & -15.85 &  8.4 &   409 & 0.101 &  0.11 &  0.06 & 1,1 \\
 71 & NGC 2527      & 246.08 &   1.85 &  8.7 &   601 & 0.038 &  0.44 & -0.09 & 1,8 \\
 72 & NGC 2539      & 233.73 &  11.12 &  9.3 &  1363 & 0.082 &  0.37 &  0.13 & 1,1 \\
 73 & NGC 2546      &  254.9 &  -1.98 &  8.7 &   919 & 0.134 &  0.07 &  0.12 & 1,1 \\
 74 & NGC 2547      & 264.45 &  -8.53 &  8.5 &   455 & 0.041 &  0.03 & -0.16 & 1,1 \\
 75 & NGC 2548      & 227.93 &  15.39 &  9.0 &   769 & 0.031 &  0.36 &  0.08 & 1,1 \\
 76 & NGC 2567      & 249.79 &   2.94 &  9.2 &  1677 & 0.128 &  0.29 & -0.09 & 1,8 \\
 77 & NGC 2571      & 249.09 &   3.54 &  9.0 &  1342 & 0.137 &  0.03 &  0.05 & 1,9 \\
 78 & NGC 2632      & 205.92 &  32.48 &  8.6 &   187 & 0.009 &  0.72 &  0.14 & 1,1 \\
 79 & NGC 2660      & 265.85 &  -3.03 &  9.1 &  2826 & 0.313 &  1.07 & -0.18 & 1,1 \\
 80 & NGC 2682      & 215.66 &  31.91 &  9.1 &   908 & 0.059 &  2.56 &  0.00 & 1,1 \\
 81 & NGC 2818      & 261.98 &   8.58 &  8.9 &  1855 & 0.121 &  0.42 & -0.17 & 1,1 \\
 82 & NGC 2972      & 274.73 &   1.75 &  8.5 &  2062 & 0.343 &  0.09 & -0.07 & 1,1 \\
 83 & NGC 3114      & 283.34 &  -3.83 &  8.3 &   911 & 0.069 &  0.12 &  0.02 & 1,1 \\
 84 & NGC 3532      & 289.55 &   1.34 &  8.3 &   486 & 0.037 &  0.31 & -0.02 & 1,1 \\
 85 & NGC 3680      & 286.78 &  16.92 &  8.2 &   938 & 0.066 &  1.19 & -0.09 & 1,1 \\
 86 & NGC 381       & 124.93 &  -1.22 &  9.2 &  1148 &   0.4 &  0.32 &  0.07 & 1,1 \\
 87 & NGC 3960      & 294.36 &   6.18 &  7.8 &  2258 & 0.302 &  0.66 & -0.17 & 1,1 \\
 88 & NGC 4349      & 299.71 &   0.82 &  7.6 &  2176 & 0.384 &   0.2 & -0.12 & 1,8 \\
 89 & NGC 5138      & 307.54 &   3.53 &  7.4 &  1986 & 0.262 &  0.09 &  0.12 & 1,1 \\
 90 & NGC 5316      & 310.22 &   0.12 &  7.7 &  1215 & 0.267 &  0.15 & -0.02 & 1,8 \\
 91 & NGC 5822      &  321.7 &   3.59 &  7.8 &   917 &  0.15 &  0.66 & -0.02 & 1,1 \\
 92 & NGC 6025      & 324.54 &  -5.88 &  7.9 &   756 & 0.159 &  0.07 &  0.19 & 1,9 \\
 93 & NGC 6067      & 329.75 &  -2.21 &  7.3 &  1417 &  0.38 &  0.11 &  0.13 & 1,1 \\
 94 & NGC 6087      & 327.76 &  -5.41 &  7.7 &   891 & 0.175 &  0.09 & -0.01 & 1,1 \\
 95 & NGC 6134      & 334.91 &  -0.19 &  7.6 &   913 & 0.395 &  0.92 &  0.18 & 1,1 \\
 96 & NGC 6208      & 333.75 &  -5.76 &  7.6 &   939 &  0.21 &  1.17 & -0.03 & 1,8 \\
 97 & NGC 6253      & 335.45 &  -6.26 &  7.1 &  1510 &  0.2  &  5    &  0.36 & 1,1 \\
 98 & NGC 6259      & 341.98 &  -1.52 &  7.5 &  1031 & 0.498 &  0.21 &  0.02 & 1,1 \\
 99 & NGC 6281      & 347.73 &   1.97 &  8.0 &   479 & 0.148 &  0.31 &  0.00 & 1,8 \\
100 & NGC 6405      & 356.59 &  -0.77 &  8.0 &   487 & 0.144 &  0.09 &  0.06 & 1,1 \\
101 & NGC 6425      & 357.94 &   -1.6 &  7.7 &   778 & 0.399 &  0.02 &  0.09 & 1,8 \\
102 & NGC 6475      & 355.85 &  -4.52 &  8.2 &   301 & 0.103 &  0.29 &  0.03 & 1,8 \\
103 & NGC 6494      &   9.89 &   2.83 &  7.8 &   628 & 0.356 &  0.29 &  0.09 & 1,1 \\
104 & NGC 6633      &   36.1 &    8.3 &  8.2 &   376 & 0.182 &  0.42 &  0.00 & 1,1 \\
105 & NGC 6705      &  27.31 &  -2.78 &  6.8 &  1877 & 0.426 &  0.2  &  0.14 & 1,1 \\
106 & NGC 6716      &  15.39 &  -9.59 &  7.7 &   789 & 0.22  &  0.09 & -0.31 & 1,9 \\
107 & NGC 6791      &  69.96 &  10.91 &  8.5 &  5853 & 0.117 &  4.39 &  0.15 & 1,1 \\
108 & NGC 6819      &  73.97 &   8.48 &  8.1 &  2360 & 0.238 &  1.49 &  0.07 & 1,1 \\
109 & NGC 6939      &   95.9 &   12.3 &  8.7 &  1185 & 0.32  &  2.21 &  0.02 & 1,1 \\
110 & NGC 6940      &   69.9 &  -7.14 &  8.2 &   770 & 0.214 &  0.72 &  0.01 & 1,1 \\
111 & NGC 7082      &  91.19 &   -2.9 &  8.6 &  1442 & 0.237 &  0.17 & -0.01 & 1,1 \\
112 & NGC 7142      & 105.34 &   9.48 &  9.0 &  1686 & 0.397 &  1.88 &  0.04 & 1,1 \\
113 & NGC 7209      &  95.49 &  -7.33 &  8.6 &  1168 & 0.168 &  0.41 & -0.12 & 1,8 \\
114 & NGC 752       & 137.18 & -23.35 &  8.8 &   457 & 0.034 &  1.12 & -0.08 & 1,1 \\
115 & NGC 7789      & 115.48 &  -5.37 &  9.7 &  2337 & 0.217 &  1.71 & -0.08 & 1,1 \\
116 & Pismis 4      & 262.86 &  -2.43 &  8.5 &   593 & 0.013 &  0.03 & -0.20 & 1,8 \\
117 & Ruprecht 18   & 239.92 &  -4.94 &  9.0 &  1056 & 0.7   &  0.04 & -0.01 & 1,1 \\
118 & Ruprecht 46   & 238.36 &    5.9 &  8.9 &   752 & 0.07  &  3.98 & -0.04 & 1,1  \\
119 & Tombaugh 2    & 232.83 &  -6.88 & 13.3 &  6300 & 0.4   &  4.0  & -0.36 & 10,11 \\
\enddata
\vspace*{2mm}

\noindent { We list 119 clusters, while in the paper, Berkley 29
was not included in our calculations.}

\tablerefs{ (1) Dias et al.2002; (2) Friel and Jane 1993; (3)
Friel 1995; (4) Edvardsson et al. 1995; (5) Gratton 2000; (6) Ann
et al. 2002; (7) Carraro, Girardi, and Marigo 2002; (8) Piatti,
Claria, and Abadi 1995; (9) Cameron 1985; (10) Kubiak et al. 1992;
(11) Brown et al. 1996. }

\end{deluxetable}

\clearpage

\begin{deluxetable}{rlrrrrrrrrrrrrrrrr}
\tabletypesize{\scriptsize}
\tablecaption{Kinematic Parameters of Open Clusters}
\tablehead{ \colhead{No.} & \colhead{ID}  &
\colhead{l} & \colhead{b}  & \colhead{V$_{r}$} & \colhead{ref} &
\colhead{$\mu_{\alpha}cos\delta$} & \colhead{$\mu_{\delta}$} &
\colhead{ref} & \colhead{R$_{GC}$} & \colhead{R$_{SUN}$} &
\colhead{$\Pi$} & \colhead{$\Theta$} & \colhead{$W$} & \colhead{$V$} &
\colhead{$\sigma_{v}$} & \colhead{age} & \colhead{[Fe/H]} }
\startdata

\enddata
\tablerefs{ (1) Rastorguev et al.(1999); (2) Baumgardt et
al.(2000); (3) Dias et al.(2002); (4) Mermilliod et al.(1996); (5)
Friel(1993); (6) Lyng\aa (1987); (7) Soubiran (2000); (8)
Mermilliod \& Mayor(1990); (9) Mermilliod \& Mayor(1989); (10)
Mermilliod et al. (1995); (11) Mermilliod et al.(1987); (12)
Clari\'a \& Mermilliod(1992);(13) Scott et al.(1995); (14) Raboud
\& Mermilliod(1998); (15) Robichon et al.(1999). }
\tablecomments{This table is available only on-line as a
machine-readable table. }
\end{deluxetable}


\begin{thebibliography}{}

\bibitem[Allen et al.(1998)]{all98} Allen, C., Carigi, L., \& Peimbert, M. 1998,
         \apj, 494, 247
\bibitem[Andrievsky et al.(2002)]{and02} Andrievsky, S. M., Kovtyukh, V. V.,
         Luck, R. E., L\`epine, J. R.D.,  Bersier, D., et al. 2002, \aap, 381, 32
\bibitem[Ann et al.(2002)]{ann02}Ann, H. B., Lee, S. H., Sung, H., Lee, M. G.,
         Kim, S. -L. et al. 2002, \aj, 123, 905
\bibitem[Baumgardt et al.(2000)]{bau00} Baumgartd, H., Dettbarn, C., \& Wielen, R.
         2000, \aap, 146, 251
\bibitem[Boissier and Prantzos(1999)]{boi99} Boissier, S., \& Prantzos, N. 1999,
         \mnras, 307, 857
\bibitem[Brown et al.(1996)]{bro96} Brown, J. A., Wallerstein, G., Geisler, D.,
         \& Oke, J. B. 1996, \aj, 112, 1551
\bibitem[Cameron(1985)]{cam85} Cameron, L. M. 1985, \aap, 147, 47
\bibitem[Carney et al.(1990)]{car90} Carney, B. W., Latham, D. W., \& Laird, J. B.
         1990, \aj, 99, 572
\bibitem[Carraro et al.(1998)]{car98} Carraro, G., Ng, Y. K.. \& Portinari, L.
         1998, \mnras, 296, 1045
\bibitem[Carraro et al.(1999)]{car99} Carraro, G., Vallenari, A., Girardi, L.,
         \& Richichi, A. 1999, \aap, 343, 825
\bibitem[Carraro et al.(2002)]{car02} Carraro, G., Girardi, L., \& Marigo, P.
         2002, \mnras, 332, 705
\bibitem[Chang et al.(1999)]{cha99} Chang, R. X., Hou, J. L., Shu, C.G. \& Fu, C. Q.
         1999, \aap, 350, 38
\bibitem[Chang et al.(2002)]{cha02} Chang, R. X., Shu, C.G., \& Hou, J.L.
         2002, \cjaa, 2, 226
\bibitem[Chen et al.(2000)]{che00} Chen, Y. Q., Nissen, P. E., \& Zhao, G. 2000,
         \aaps, 141,491
\bibitem[Chiappini et al.(1997)]{chi97} Chiappini, C., Matteucci, F., \& Gratton, R.
         1997, \apj, 477, 765
\bibitem[Chiappini et al.(2001)]{chi01} Chiappini, C., Matteucci, F., \& Romano, D.
         2001, \apj, 550, 1044
\bibitem[Clari\'a and Mermilliod(1992)]{cla92} Clari\'a, J.J., \& Mermilliod, J.-C.
         1992, \aapss, 95, 429
\bibitem[Corder and Twarog(2001)]{cor01} Corder, S., \& Twarog, B. A. 2001,
         \aj, 122, 895
\bibitem[Cui et al.(2000)]{cui00} Cui, C. Z., Zhao, G., Zhao, Y. H., \& Shi, J. R.
         2000, \sinc (series A), 30, 953
\bibitem[Dias et al.(2001)]{dia01} Dias, W. S., L\'epine, J. R., \& Alessi, B. S.
         2001, \aap, 376, 441
\bibitem[Dias et al.(2002)]{dia02} Dias, W. S., Alessi, B. S., Moitinho, A., \&
         L\'epine, J. R. 2002, \aap, 389, 871
\bibitem[Edvardsson et al.(1993)]{edv93} Edvardsson, B., Andersen, J.,
         Gustafsson, B., et al. 1993, \aap, 275, 101
\bibitem[Edvardsson et al.(1995)]{edv95} Edvardsson B., Pettersson B.,
         Kharrazi M., et al. 1995, \aap, 293, 75
\bibitem[Eggen, Lynden-Bell and Sandage(1962)]{els62} Eggen, O. J., Lynden-Bell, D.,
         \& Sandage, A. R. 1962, \apj, 136, 748
\bibitem[ESA(1997)]{esa97} ESA, 1997, The Hipparcos and Tycho Catalogues
         (European Space Agency (ESA))
\bibitem[Feltzing et al.(2001)]{fel01} Feltzing, S., Holmberg, J. \& Hurley, J.R.
         2001, \aap, 377, 911
\bibitem[Friel(1989)]{fri89} Friel, E. D. 1993, \pasp, 101, 244
\bibitem[Friel(1995)]{fri95} Friel, E. D. 1995, \araa, 33, 381(F95)
\bibitem[Friel(1999)]{fri99} Friel, E. D. 1999, \apss, 265, 271
\bibitem[Friel and Janes(1993)]{fri93} Friel, E. D., \& Janes, K. A. 1993,
         \aap, 267, 75 (FJ93)
\bibitem[Gratton(2000)]{gra00} Gratton, R. 2000, ASPC, 198, 225
\bibitem[Carney and Harris(2001)]{car01} Carney, B. W., and Harris, W. E. 2001,
         Star Clusters: Saas-Fee Advanced Course 28, ed. L.Labhardt and B.Binggeli
         (Springer, Swiss Society for Astrophysics and Astronomy)
\bibitem[Henry and Worthey(1999)]{hen99}Henry, R. B. C., \& Worthey, G. 1999,
         \pasp, 111, 919
\bibitem[H$\o$g et al.(2000)]{hog00} H$\o$g, E., Fabricius, C., Makarov, V. V.,
         Urban, S., Corbin, T., et al. 2000, \aap, 355, L27
\bibitem[Hou et al.(2000)]{hou00} Hou, J. L., Prantzos, N., \& Boissier, S.
         2000, \aap, 362, 921
\bibitem[Hou et al.(2002)]{hou02} Hou, J. L., Chang, R.X., \& Chen, L.
         2002, \cjaa, 2,17
\bibitem[Janes(1979)]{jan79} Janes, K. A. 1979, \apjs, 39, 135
\bibitem[Janes et al.(1988)]{jan88} Janes, K. A., Tilley, C.. \& Lyng\aa, G.
         1988, \aj, 95, 771
\bibitem[Janes and Phelps(1994)]{jan94} Janes, K. A., \& Phelps, R. L. 1994, \aj,
         108, 1773
\bibitem[Kaluzny(1994)]{kal94} Kaluzny, J. 1994, A\&AS, 108, 151
\bibitem[Kubiak et al.(1992)]{kub92} Kubiak, M., Kaluzny, J., Krzeminski, W., \&
         Mateo, Mario, Acta Astronomica, 1992, 42, 155
\bibitem[Larsen(2002)]{lar02} Larsen, S. S. 2002, IAU Symposium series, 207
         (in press)
\bibitem[Lyng\aa(1987)]{Lyn87} Lyng\aa, G. 1987, Computer Based Catalogue of
         Open Cluster Data, 5th ed. (Strasbourg: CDS)
\bibitem[Maciel and K\"{o}ppen(1994)]{mac94} Maciel, W. J., \& K\"oppen, J.
         1994, \aap, 282, 436
\bibitem[Maciel and Quireza(1999)]{mac99} Maciel, W. J., \& Quireza, C.
         1999, \aap, 345, 629
\bibitem[Maciel and Costa(2002)]{ma02a} Maciel, W. J., \& Costa, R. D. D.
         2002, IAU Symp. 209, in press (astro-ph/0112210)
\bibitem[Maciel et al.(2002)]{ma02b} Maciel, W. J., Costa, R. D. D.,
         \& Uchida, M. M. M. 2002, \aap, accepted (astro-ph/0210470)
\bibitem[Mermilliod et al.(1995)]{mer95} Mermilliod, J.-C., Andersen, J.,
         Nordstr\"{o}m, B., \& Mayor, M.  1995, \aap, 299, 53
\bibitem[Mermilliod et al.(1987)]{mer87} Mermilliod, J.-C., Mayor, M., \&
         Burki, G. 1987, A\&AS, 70, 389
\bibitem[Mermilliod et al.(1996)]{mer96} Mermilliod, J.-C., Hustamendia, G.,
         del Rio, G., \& Mayor M.  1996, \aap, 307, 80
\bibitem[Mermilliod and Mayor(1989)]{mer89}  Mermilliod, J.-C., \& Mayor, M.
         1989, \aap, 219, 125
\bibitem[Mermilliod and Mayor(1990)]{mer90}  Mermilliod, J.-C., \& Mayor, M.
         1990, \aap, 237, 61
\bibitem[Meusinger et al.(1991)]{meu91} Meusinger, H., Stecklum, B., \&
         Reimann, H.-G. 1991, \aap, 245, 57
\bibitem[Moll\`a et al.(1997)]{mol97} Moll\`a, M., Ferrini, F., \& Diaz, A. I.
         1997, \apj, 475, 519
\bibitem[Noriega-Mendoza and Ruelas-Mayorgo(1997)]{nor97} Noriega-Mendoza, H.,
         \& Ruelas-Mayorgo, A. 1997, \aj, 113, 722
\bibitem[Ojha et al.(1996)]{ojh96} Ojha, D. K., Bienayme, O., Robin, A. C.,
         Creze, M., \& Mohan, V. 1996, \aap, 311, 4560
\bibitem[Panagia and Tosi(1981)]{pan81} Panagia, N., \& Tosi, M. 1981,
         \aap, 96, 306
\bibitem[Phelps et al.(1994)]{phe94} Phelps, R. L., Janes, K. A., \&
         Montgomery, K. A. 1994, \aj, 107, 1079
\bibitem[Piatti et al.(1995)]{pia95} Piatti, A., Claria, J. J., \& Abadi, M. G.
         1995, \aj, 110, 2813
\bibitem[Piatti et al.(1998)]{pia98} Piatti, A. E., Clari\'a, J. J., Bica, E.,
         Geisler, D., \& Minniti, D. 1998, \aj, 116, 801
\bibitem[Portinari and Chiosi(2000)]{por00} Portinari, L., \& Chiosi, C. 2000,
         \aap, 355, 929
\bibitem[Prantzos and Aubert(1995)]{pra95} Prantzos, N., \& Aubert, O. 1995,
         \aap, 301, 69
\bibitem[Raboud and Mermilliod(1998)]{rab98} Raboud, D., \& Mermilliod, J.-C.
         1998, \aap, 329, 101
\bibitem[Rastorguev et al.(1999)]{ras99} Rastorguev, A. S., Glushkova, E. V.,
         Dambis, A. K., \& Zabolotskikh, M. V. 1999, AZh Pisma, 25, 689
         (English transl. Astron. Lett., 25, 595)
\bibitem[Robichon et al.(1999)]{rob99} Robichon, N., Arenou, F., Mermilliod, J.-C.,
         \& Turon, C. 1999, \aap, 345, 471
\bibitem[Samland et al.(1997)]{sam97} Samland, M., Hensler, G., \& Theis, Ch.
         1997, \apj, 476, 544
\bibitem[Scott et al.(1995)]{sco95} Scott, J. E., Friel, E. D., \& Janes, K. A.
         1995, \aj, 109, 1706
\bibitem[Simpson et al.(1995)]{sim95} Simpson, J. P., Colgan, S. W. J.,
         Rubin, R. H., Erickson, E. F., \& Haas, M. R. 1995, \apj, 444, 721
\bibitem[Soubiran et al.(2000)]{sou00} Soubiran, C., Odenkirchen, M.,
         \& Le Campion, J.-F. 2000, \aap, 357, 484
\bibitem[Tosi(1988)]{tos88} Tosi, M. 1988, \aap, 197, 47
\bibitem[Twarog(1980)]{twa80} Twarog, B. A. 1980, \apj, 242, 242
\bibitem[Twarog(2002)]{twa02} Twarog, B. A. 2002, private communication
\bibitem[Twarog Ashman and Anthony-Twarog(1997)]{twa97} Twarog, B. A.,
         Ashman, K. A., \& Anthony-Twarog, B. J. 1997, \aj, 114, 2556 (TAA97)
\bibitem[van der Bergh and MMcClure(1980)]{van80} van der Bergh, S., \&
         McClure, R.D. 1980, \aap, 88, 360
\bibitem[Wyse(2001)]{wys01} Wyse, R. F. G. 2001, ASPC, 230, 71

\end{thebibliography}
\end{document}